# Structural and dielectric properties of Pb$_{(1-x)}$(Na$_{0.5}$Sm$_{0.5}$)$_x$TiO$_3$ ceramics

Arun Kumar Yadav[1], Anita[1], Sunil Kumar[1], V. Raghavendra Reddy[3], Parasharam M. Shirage[1,2], Sajal Biring[4], Somaditya Sen[1,2]*

[1] Discipline of Metallurgy Engineering and Materials Science, Indian Institute of Technology, Khandwa Road, Indore-453552 India
[2] Discipline of Physics, Indian Institute of Technology, Khandwa Road, Indore-453552 India
[3] UGC-DAE Consortium for Scientific Research, University Campus, Khandwa Road, Indore-452001, India
[4] Electronic Engg., Ming Chi University of Technology, New Tapei City, Taiwan

Corresponding Author: sens@iiti.ac.in

## Abstract:

A correlation between structure and vibrational properties related to a ferroelectric to paraelectric phase transition in perovskite Pb$_{(1-x)}$(Na$_{0.5}$Sm$_{0.5}$)$_x$TiO$_3$ (0≤$x$≤0.5) (PNST-$x$) polycrystalline powders is discussed. Substitution leads to reduction of tetragonality which is associated with a shift of the phase transition to lower temperatures. The nature of the phase transition gets diffused with increasing substitution.

## Introduction

PbTiO$_3$ is known to be a strong ferroelectric perovskite compound. A structural phase transition from tetragonal (*P4mm*) to cubic (*Pm3m*) phase at 763K is observed in PbTiO$_3$. Tetragonality, measured by *c/a* ratio is frequently related to ferroelectric properties. For PbTiO$_3$, c/a ratio is ~1.06 at room temperature [1-5]. Properly substituted PbTiO$_3$ has different applications as capacitors, nonvolatile memories, ultrasonic transducers, pyroelectric infrared sensors, piezoelectric actuators due to their highly anisotropic nature of elastic, piezoelectric, and electro-optic properties etc. [6-8].

Physical properties and device applications of ferroelectric perovskite (ABO$_3$) compounds like PbTiO$_3$ and BaTiO$_3$ have been studied extensively due to their technological importance. Based on the fact that PbTiO$_3$ is toxic and volatile in nature and not environmental friendly many reduced lead content ferroelectric materials has been investigated, such as Pb(Mg$_{1/3}$Nb$_{2/3}$)O$_3$-PbTiO$_3$, Ba(Mg$_{1/3}$Nb$_{2/3}$)O$_3$−$x$PbTiO$_3$, Ba(Zn$_{1/3}$Nb$_{2/3}$)O$_3$−$x$PbTiO$_3$, Ba(Yb$_{1/2}$Nb$_{1/2}$)O$_3$−$x$PbTiO$_3$, Ba(Sc$_{1/2}$Nb$_{1/2}$)O$_3$−$x$PbTiO$_3$ and BaSnO$_3$−$x$PbTiO$_3$ [9-12]. Isovalent or heterovalent substitution on Pb$^{2+}$ site, lattice anisotropy is reduced.

Diffuse phase transition in dielectric response has a broad peak with temperature rather than sharp peak in normal ferroelectric. The diffusion of the peak comes from the fact that the phase transition in different micro regions takes place at different temperatures. The signature of this is typically due to heterovalent disorder substitution [13-14].

Motivation of our research is to explore the effect of substitution of *Pb* by *Na/Sm* on the structural, vibrational and dielectric properties of the newly synthesized compounds of $Pb_{(1-x)}(Na_{0.5}Sm_{0.5})_xTiO_3$ with ($0 \leq x \leq 0.5$). Although $PbTiO_3$ is an extensively studied material, however, hardly any report is available in the literature on $Pb_{(1-x)}(Na_{0.5}Sm_{0.5})_xTiO_3$ compounds.

**Synthesis**

Polycrystalline powders of $Pb_{(1-x)}(Na_{0.5}Sm_{0.5})_xTiO_3$ ($0 \leq x \leq 0.5$) ceramics were prepared using sol-gel process. Precursors were used to synthesize these materials, Lead (II) nitrate (99.999%, Alfa Aesar), Samarium oxide (99.999%, Alfa Aesar), Sodium Nitarte and Dihydroxybis (ammonium lactate) titanium (IV), 50% w/w aqua solution (Alfa Aesar). The stoichiometric solutions of each precursor were prepared with double distilled water in a separate beaker. Samarium oxide is soluble in dilute nitric acid and it was added to the titanium solution followed by the addition of the sodium and lead solution. A solution of citric acid and ethylene glycol of 1:1 molar ratio was prepared in a separate beaker as gel former and was thereafter added after vigorous stirring and heating (~70 $^o$C) on hot plate. After mixing of all precursors in a beaker we added ammonium hydroxide to maintain the pH value of the solution at 8. The solution was continuously stirred and heated (~70 $^o$C) on the hot plate to form gel. After that, gels were burnt into a big beaker of two liter inside a fume hood. Burnt powders were heated at 450 $^o$C for 12 h as a denitrification of the powder. Carefully grinding of these powders were mixed with 5% PVA solution as a binder and uniaxially pressed into discs of 13 mm diameter and 1.5 mm thickness. These pellets were sintered at 600 $^o$C for 6 h to burn out the binder continued with 1050 $^o$C for 4h.

X-ray powder diffraction was performed using a Bruker D2 Phaser X-ray Diffractometer to ensure the phase purity of the sintered samples. The HR Raman spectrometer was done with Czerny-Turner type achromatic spectrograph with spectral resolution of 0.4 cm$^{-1}$/pixel and the source of excitation is 632.8 nm. Microstructure and grain size of the sintered pellets were investigated by Supra55 Zeiss field emission scanning electron microscope.

For electrical property measurements, sintered pellets with relative density around 90-92% of the theoretical value were prepared. Electrodes were prepared using silver paste painted on both side of the sintered pellet. The silver coated pellets were cured at 550$^o$C for 10 minutes. To avoid any moisture to get adhered to the sample, we heated the samples at 200 $^o$C for 10 minutes. Dielectric response was measured using a Newtons 4$^{th}$ LTD PSM 1735 phase sensitive with a signal strength of 0.5 V. Ferroelectric (*P-E*) studies were carried out by ferroelectric loop (*P-E*) tracer (M/s Radiant Instruments, USA). During the ferroelectric measurements, the samples were immersed in silicone oil to prevent electric arcing, at high applied voltages.

**Results and discussion**

X-ray powder diffraction (XRD) patterns of the PNST-$x$ powders with $0 \leq x \leq 0.5$, were analyzed (Fig. 1). A tetragonal *P4mm* phase is detected for compositions $x$ = 0, 0.10, 0.20, 0.30 and 0.40 and cubic *Pm3m* phase for $x$=0.5. There are two extra peaks of trace amount of $Sm_2TiO_5$ at 27.6° and 29.3°. This $Sm_2TiO_5$ phase may be due to volatile nature of *Pb*. A LeBail profile fitting of the XRD data was done with Topas3.2 software to calculate lattice parameters. With increasing substitution, the separation between the (001) and (100) peaks reduces and thereafter merges at $x$=0.5. That *Na/Sm* is substituting lead is indicated from the continuous decrease in the volume of the substituted compounds, due to decrease of lattice parameter '*c*', while '*a*' is nearly constant. The average ionic radii of A-site are calculated using the relation [15]:

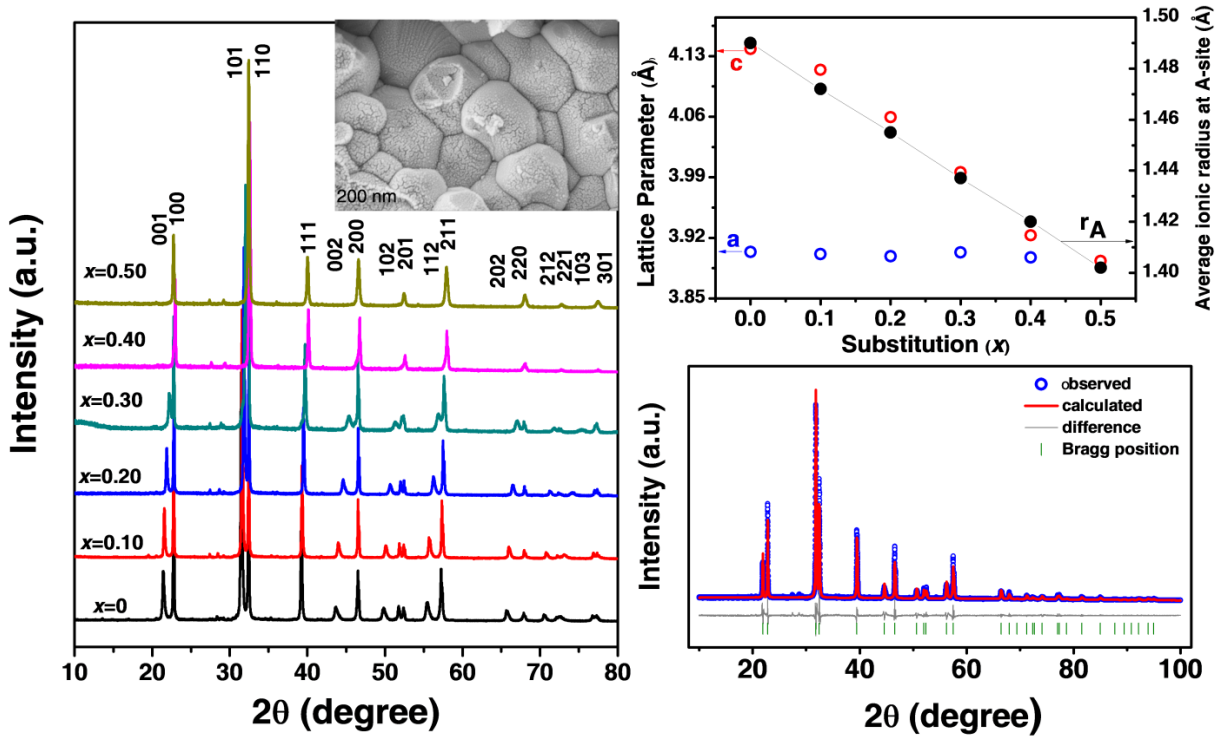

**Fig. 1.** (a) X-ray diffraction patterns of the compound $Pb_{(1-x)}(Na_{0.5}Sm_{0.5})_xTiO_3$ ($0 \leq x \leq 0.5$) (b) Lattice parameter and average ionic radius at A-site versus substitutions $x$ (c) LeBail profile fitting for $x$=0.20 with tetragonal space group *P4mm*.

$$r_A = [(1-x)r_{Pb^{2+}}] + [\tfrac{x}{2}(r_{Sm^{3+}} + r_{Na^+})] \quad (1)$$

where, Shannon's radii values are given as, $r_{Pb^{2+}} = 1.49$ Å, $r_{Sm^{3+}} = 1.24$ Å, and $r_{Na^+} = 1.39$ Å. Note that $r_A$ decreases linearly with substitution [Table 1]; $r_A$ also determines the composition dependence of the tolerance factor ($t$), calculated by,

$$t = \frac{r_A + r_O}{\sqrt{2}(r_B + r_O)} \quad (2)$$

where, $r_A$, $r_B$ and $r_O$ are the effective radii of the A-site cation, B-site cation and the oxygen ion respectively. Tolerance factor is a quantitative measure of the mismatch between the bonding requirements of the A-site and B-site cations in the perovskite $ABO_3$ and subsequently reflects the structural distortion such as rotation and tilt of the octahedral. Since the substituent at A-site ionic radii are lesser compare to *Pb* so tilt as well as non-centrosymmetric distortion will reduce. It is also reflected with the merging of peaks in XRD that indicate from non-centrosymmetric to centrosymmetric structure. Again we need to mention that for tolerance factor ~1, a perovskite phase is expected [16-17]. The tolerance factor of the PNST-*x* samples decreases from 1.019 as in *x*=0 to 0.988 for *x*=0.5 [table 1]. Hence, the general tendency of these structures is to be in the perovskite phase as t ~ 1. Also notice the strong similarity between the nature of decrement of the *c*-axis with the reduction in effective ionic radius of the A-site although *a*-axis is invariant. This indicates that *Na/Sm* substitution does not influence the *a*-axis but varies proportionately the *c*-axis. Microstructure studies were performed of PNST-*x* compounds. The average grain size with standard deviation decreases from 1.03 ± 0.36 μm as in *x*=0.10 to 0.60 ± 0.17 μm for *x*=0.50 with substitution are reported in table. 1. Microstructure of fracture surface of *x*=0.40 was shown in Fig. 1a.

There is a subtle balance between the long-range Coulomb interaction and short-range forces in materials. Domain structure and defect determine the long range Coulomb interaction which in turn makes the ferroelectric transitions highly sensitive. Such changes lead to the splitting of longitudinal optical (*LO*) and transverse optical (*TO*) phonons [18-19]. Hence Raman spectroscopy is an excellent tool in qualitatively assessing retention of domain structure, defects and structural distortions thereby understand deformations and lattice strains associated with substitution in ferroelectrics like in PNST-*x*.

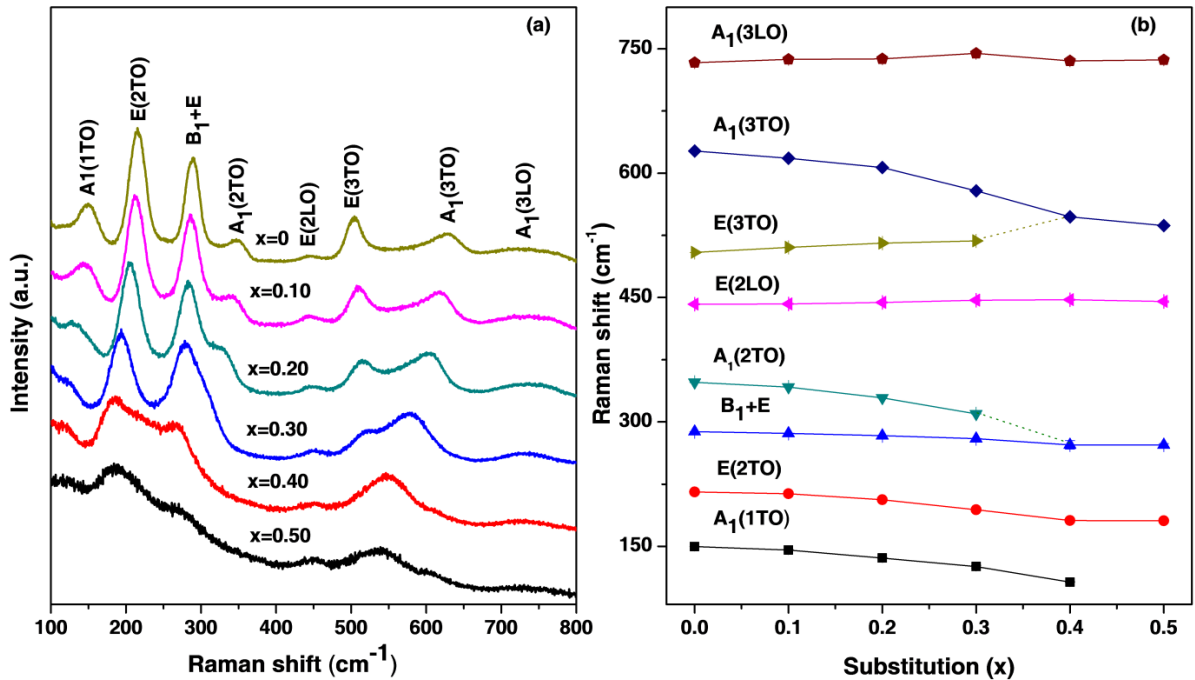

**Fig. 2.** (a) HR Raman spectroscopic measurement of the compounds $Pb_{(1-x)}(Na_{0.5}Sm_{0.5})_xTiO_3$ $0 \leq x \leq 0.5$. (b) Variations of mode after fitting the Raman data.

The Raman spectra of PNST-$x$ are shown in Fig. 2a. All the phonon vibrations of these compounds correspond to known vibrational modes of pure $PbTiO_3$. Intensities decreased, peak widths increased and positions varied for different modes with substitution. The cubic paraelectric phase of pure $PbTiO_3$ belongs to the space group with *12* optic modes at the point of the Brillouin zone in the $3T_{1u}+T_{2u}$ representation of the $O_h$ point group. The $3T_{1u}$ modes are Raman inactive but infrared active while $T_{2u}$ modes (silent modes) are both Raman and infrared inactive. Therefore the cubic phase has no Raman active modes [20-22]. Although in XRD we see that the lattice is becoming cubic, the very fact that in our samples Raman modes is visible up to $x=0.50$, hints at retention of ferroelectricity until $x=0.50$. After fitting the Raman data, energy variations of the phonon modes with composition are observed [Fig. 2b]. Except $A_1(3LO)$, $E(3TO)$ and $E(2TO)$ which are blue shifted, all other modes are red shifted.

The $A_1(1TO)$ transverse optical mode relates to the relatively opposite vibrations of the *O-B-O* chains to the *Pb* sub lattice along *c*-axis in the $ABO_3$ perovskite structure. It is also dependent on the displacement of the $BO_6$ octahedron relative to *Pb* atoms [23-24]. It has a direct relation to the order parameter. $A_1(1TO)$ is generally called "soft mode" due to losing energy as the sample undergoes a tetragonal to cubic phase transition [25-26]. In PNST-$x$, $A_1(1TO)$ mode loses intensity and energy with increasing substitution indicating a direct relation to our XRD results. Reduction in the energy is a direct consequence of the reduction in the *c*-axis in spite of reduction of effective mass of the A-site. Such a reduction may be due to increase of lattice distortions due to *Na/Sm* introduction. Please note that the energy of $A_1(3TO)$ reduces while $E(3TO)$ modes marginally increases to merge in $x=0.4$ sample. The relative vibrational motion of the *Ti* atoms with respect to the oxygen atoms along the *O-Ti-O* chains along *c*-axis result in $A_1(3TO)$ soft modes. As TO modes vibrate along the direction of the spontaneous polarization (*c*-axis) in $PbTiO_3$, the $A_1(3TO)$ mode is very important. The $E(3TO)$ mode is one of the modes which has gained energy. This mode is a result of vibrations of *Ti* and planar *O* ions with respect to each other along the '*a*' and '*b*' axes. Please note, in parent $PbTiO_3$, the displaced off-centered *Ti* ion is at a strained bonding with the planar oxygen. With the reduction in tetragonality, these *Ti* ions will be nearer to the *O*-plane and the structure will become more centro-symmetric. Thereby the strain in the *O-Ti-O* bond will be reduced and the ions will be capable of vibrating more energetically. In congruence with our XRD results the reduction of tetragonality is also reflected in our Raman results.

Dielectric measurement is a powerful tool to study the effect of *Na/Sm* substitution on the phase transition of the PNST-$x$ ceramics. A phase pure, sufficiently dense and mechanically robust $PbTiO_3$ pellet was extremely difficult to fabricate and hence dielectric property of $PbTiO_3$ phase was not investigated. With substitution preparation of pellets, appropriate for dielectric measurements, became easier and dielectric properties including dielectric constant, impedance, capacitance and dielectric loss factor were measured in the temperature range 300-750 K for various frequencies in 10 Hz – 1 MHz range. Here we will concentrate on the dielectric constant only to discuss a probable ferroelectric to paraelectric phase transition of these materials.

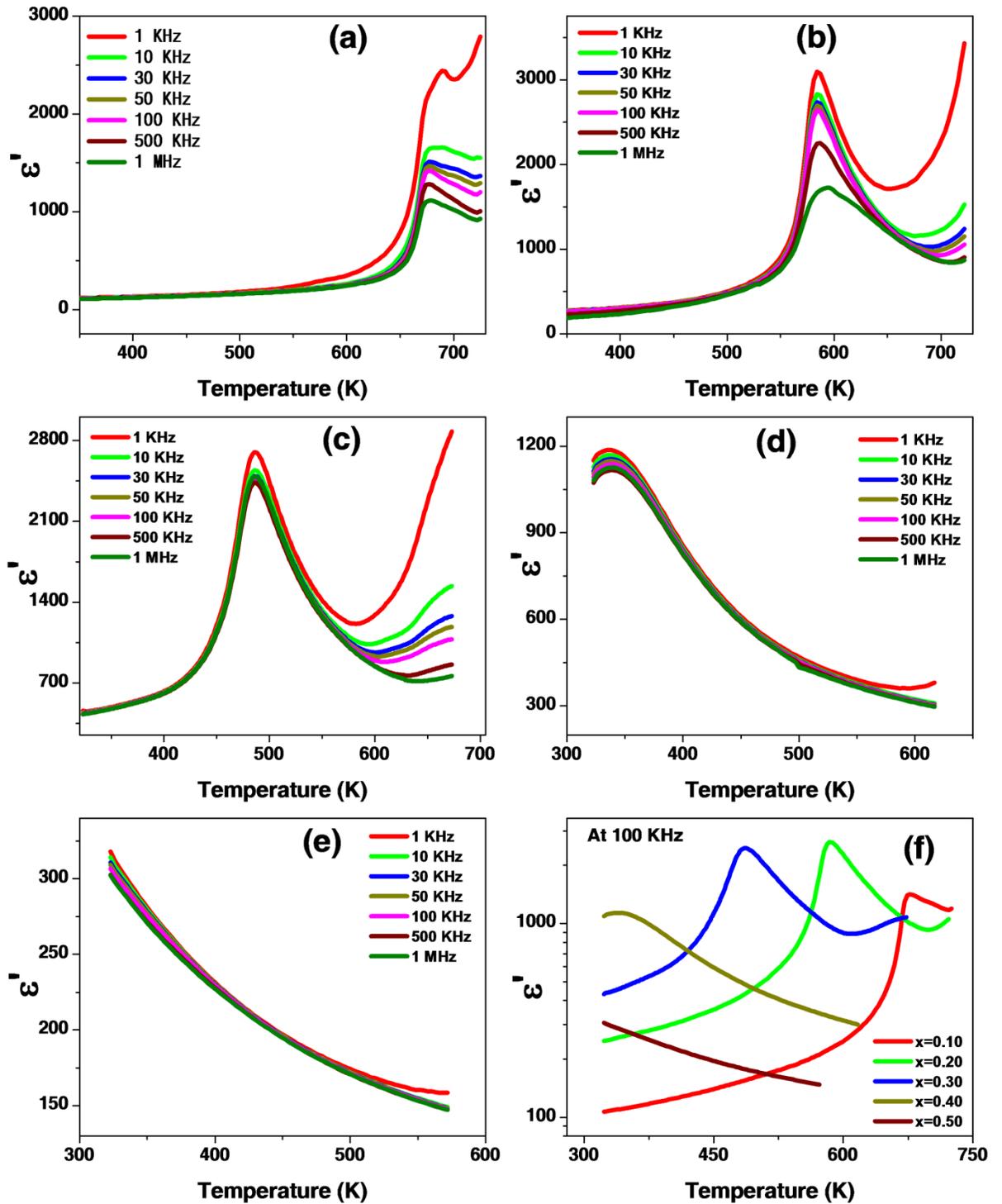

**Fig. 3.** Temperature dependent dielectric constant of $Pb_{(1-x)}(Na_{0.5}Sm_{0.5})_xTiO_3$ for composition (a) $x=0.10$, (b) $x=0.20$, (c) $x=0.30$, (d) $x=0.40$, (e) $x=0.50$, (f) Dielectric constant vs temperature at 100 kHz for $0.10 \leq x \leq 0.50$ composition.

Temperature dependent dielectric constant [Fig. 3(a-e)] for different fixed frequencies as a function of temperature for PNST-$x$ ceramics shows peaks above room temperature for all samples except for $x=0.5$. The temperature of dielectric maximum, $T_m$, decreases with increasing substitution from ~677 K in $x=0.1$ to ~338 K in $x=0.4$ for all frequencies. At 1 kHz, $\varepsilon'$-$T$ peak around 670 K seems to be overshadowed by some other phenomenon. This may be a space-charge polarization effect. Values of dielectric constant at $T_m$ for $x=0.1$ show significant frequency dispersion. The $x=0.5$ sample shows a gradually decreasing $\varepsilon'$ with increase in temperature in the entire range suggesting $T_m$ below room temperature. It can be further noticed that the frequency dispersion in dielectric constant around $T_m$ decreases with increasing $Na/Sm$ content. Such frequency dispersion trend can be explained by the incidence of space charge polarization which becomes significant at low frequencies and high temperatures [27].

Pure $PbTiO_3$ is known to have a sharp ferroelectric to paraelectric transition around 763 K (Curie temperature) accompanied by a structural transition from polar tetragonal to non-polar cubic phase [28]. This structural phase transition manifests as a sharp and frequency independent peak in dielectric permittivity versus temperature plot. Fig. 3f shows $\varepsilon'$ as a function of temperature measured at 100 kHz for various PNST-$x$ samples. It can be clearly seen that the temperature of dielectric maximum decreases linearly with increasing $x$. $Pb$ $6s^2$ lone pair electrons help in stabilizing the large tetragonal strain (~ 6%) in $PbTiO_3$. Our XRD analysis confirmed that $Na^+/Sm^{3+}$ substitution at $Pb^{2+}$ (with $6s^2$ lone pair) site modifies the tetragonal strain in $PbTiO_3$ and finally for $x=0.5$ exhibits a '*cubic*' structure. Note that such a structural change is also reflected in phase transition temperature, $Na/Sm$ substitution requiring lesser thermal energy for the ferroelectric-paraelectric phase transition. Similar trend has been observed in other perovskite related ferroelectrics [15, 29].

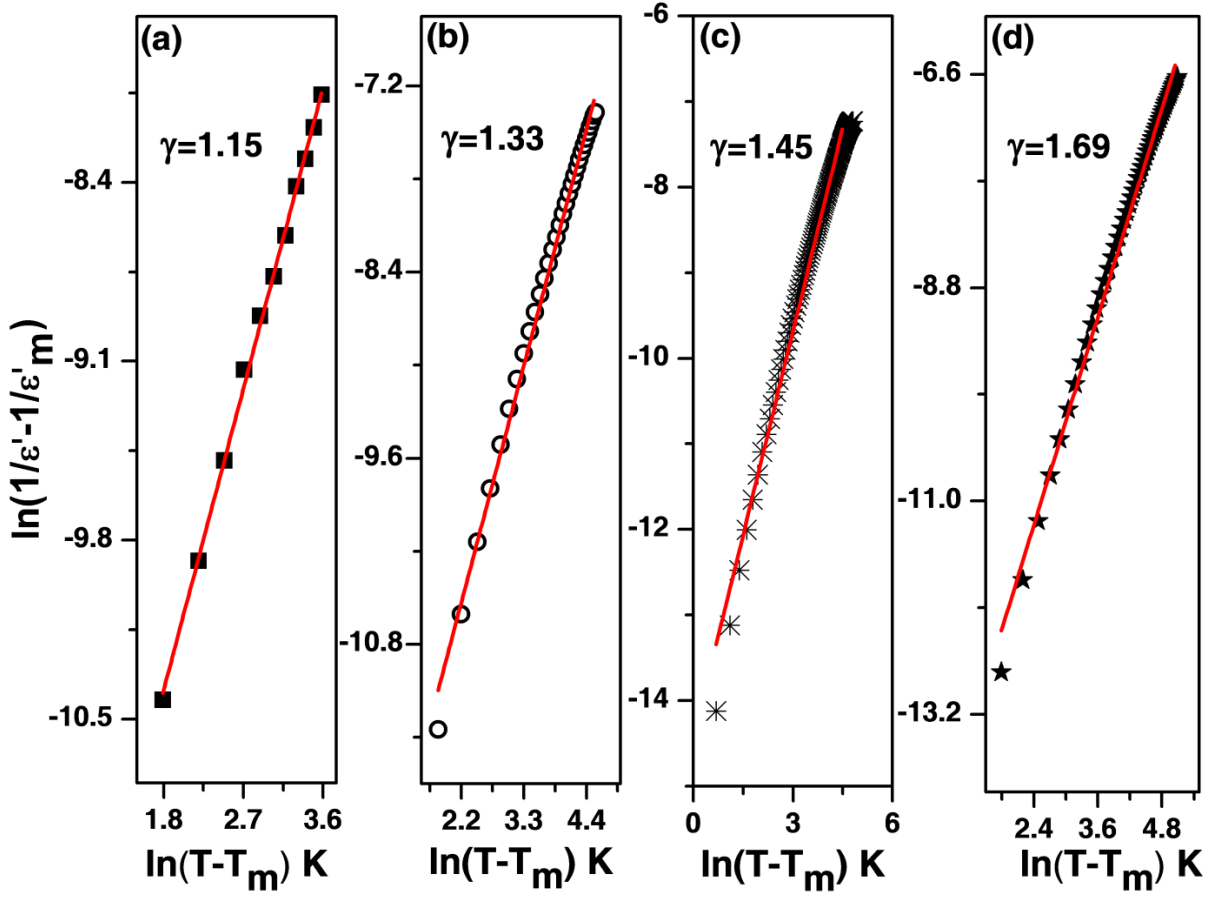

**Fig. 4.** Plot between $\ln(\frac{1}{\varepsilon'} - \frac{1}{\varepsilon'_m})$ verses $\ln(T-T_m)$ with linear fitting of the compound PNST-$x$ at 100 kHz. Where solid lines are representing the fitted data and symbols experimental data.

Diffuse phase transition is usually observed in perovskites with random distribution of different types of ions on structurally identical sites in lattice [30-31]. It must be noted that diffuse phase transition exhibit a broad instead of a sharp change of structure and properties at the Curie point in normal ferroelectric materials [32-33]; consequently, the phase transition characteristics of such materials are known to diverge from the characteristic Curie-Weiss behavior and can be described by a modified Curie-Weiss formula [34].

$$\frac{1}{\varepsilon'} - \frac{1}{\varepsilon'_m} = C^{-1}(T - T_m)^\gamma \tag{3}$$

where, $C$ is Curie-Weiss constant and the degree of diffuseness, $\gamma$ (1≤γ≤2; where γ=1 gives a sharp change while an ideal diffuse phase transition gives γ=2). An apparent increase in diffuseness of peak in $\varepsilon'$-$T$ plots is reflected in the least-square linear fitting of $\ln(\frac{1}{\varepsilon'} - \frac{1}{\varepsilon'_m})$ versus $\ln(T - T_m)$ curves [Fig. 4] at a frequency of 100 kHz of PNST-$x$ ceramics. A reasonably good fit is observed for all compositions. The degree of diffuseness, $\gamma$ was calculated from the slop of linear fit. $\gamma$ was found to increases from $1.15\pm0.02$ for $x=0.1$ to $1.69\pm0.01$ for $x=0.5$ indicating a significant increase in diffuseness of phase transition in doped samples. Compositional disorder arising due to the random distribution of $Na^+/Sm^{3+}$ seems to be responsible for observed diffuse phase transition in PNST-$x$ ceramics [33] (Fig. S1) Room temperature dielectric constant ($\varepsilon'$) of PNST-$x$ ceramics gradually increases with increasing $x$ in the range $0.10 \leq x \leq 0.40$ and thereafter decreases for $x=0.5$ (Table. 1). This behavior can be attributed to the higher polarizability of $Na$ and $Sm$ than $Pb$ [35]. The decrease in $\varepsilon'$ in $x=0.5$ is due to the paraelectric phase of the sample at room temperature whereas the other samples are in ferroelectric phase.

**Table. 1.** Room temperature dielectric constant and loss, average grain size, tolerance factor, phase transition temperature, apparent remnant polarization and coercive field for the compounds PNST-$x$.

| Substitution ($x$) | $\varepsilon'$ (100 kHz) | Tan$\delta$ (100 kHz) | Average grain size (~$\mu$m) | Tolerance factor (t) | $T_m$ (K) | $2P_r$ ($\mu C/cm^2$) | $2E_C$ ($kV/cm$) |
|---|---|---|---|---|---|---|---|
| 0.10 | 178 | 0.012 | $1.03 \pm 0.36$ | 1.019 | 677 | 0.43 | 14.82 |
| 0.20 | 236 | 0.017 | $0.88 \pm 0.25$ | 1.013 | 584 | 2.14 | 25.84 |
| 0.30 | 397 | 0.026 | $0.67 \pm 0.24$ | 1.007 | 486 | 47.65 | 43.25 |
| 0.40 | 1225 | 0.038 | $0.60 \pm 0.25$ | 1.001 | 338 | 18.56 | 16.99 |
| 0.50 | 364 | 0.073 | $0.60 \pm 0.17$ | 0.995 | …. | …. | …. |

To confirm the ferroelectricity in PNST-$x$, polarization versus electric field ($P$-$E$) hysteresis loop measurements were carried out for all the compositions. The hysteresis loops were measured at a frequency of 1 Hz at room temperature. For composition with $x=0.10$, [Fig. 5], $P$-$E$ hysteresis loops are not typical ferroelectric loops but are more like those usually obtained for lossy dielectric materials [36]. Observed apparent remnant polarization and coercive field values at an applied electric field ~ 50 $kV/cm$ for all the samples are given in Table1. $P$-$E$ hysteresis loops obtained for compositions with $x = 0.20, 0.30$, and $0.40$ are typical hysteresis loops obtained for ferroelectric ceramics while an almost linear relation between P and E is observed for composition with $x=0.50$. This suggests that sample with $x=0.50$ is macroscopically in paraelectric state at room temperature and it is expected as corroborated by the XRD, Raman and dielectric studies. Sample with composition $x=0.30$ shows the highest apparent remanent polarization of 47.65 $\mu C/cm^2$ and a coercive field of 43.25 $kV/cm$ at an applied electric field of 50 $kV/cm$.

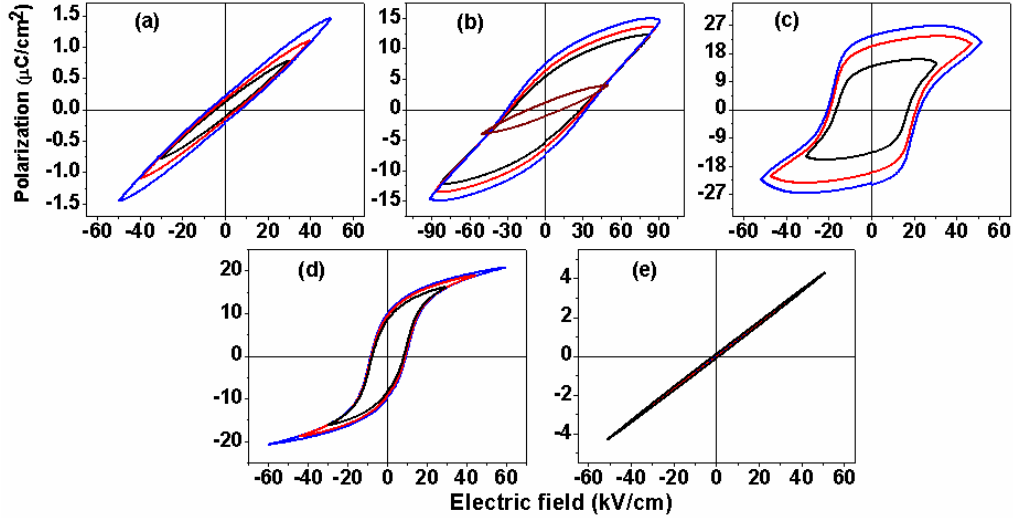

**Fig. 5.** *P-E* loop measurement of the samples PNST-*x*, (a) *x*=0.10, (b) *x*=0.20, (c) *x*=0.30, (d) *x*=0.40, and (e) *x*=0.50

From the structural point of view, composition with *x*=0.10 is expected to be ferroelectric. However, at applied electric fields up to 50-60 *kV/cm*, no signature of domain switching (or ferroelectric ordering) is discernable in *P-E* hysteresis loops. It was not possible to apply more voltage due to high leakage current in this composition [37].

Drastic improvement in room temperature ferroelectric properties in *Na/Sm* doped samples can be attributed to the cumulative effect of improved density, reduced volatilization of *Pb*, associated oxygen vacancies, increased dielectric constant and reduced tetragonal strain. It is well known that oxygen vacancies are created due to *Pb* loss during high-temperature sintering of PbTiO$_3$ based perovskite ceramics and are expressed by following Kröger-Wink equation:

$$Pb_{Pb}^{\times} + O_o^{\times} \leftrightarrow \text{PbO}(\uparrow) + V_{Pb}'' + V_o^{\cdot\cdot} \qquad (1)$$

These oxygen vacancies can hop easily due to their high mobility in applied high electric field and accumulate in the places with low free energy, such as domain walls and grain boundaries. Accumulation of these oxygen vacancies at the domain boundary causes domain pinning which restricts polarization switching [38-39].

## Conclusion

Sodium (Na) and samarium (Sm) doped PbTiO$_3$ ceramic samples PNST-*x* have been successfully synthesized by the modified Pechini sol-gel process. Phase confirmation was shown by X-ray diffraction pattern of the Pb$_{(1-x)}$(Na$_{0.5}$Sm$_{0.5}$)$_x$TiO$_3$ ($0 \leq x \leq 0.5$) samples . From HR Raman spectroscopy measurement, we investigated the variations of vibrational modes with composition and also relate with ferroelectric property. High temperature dielectric study was done to see the variations of the phase transition temperature with composition. Decrease of phase transition temperature is due to replacement of *Pb* ($6s^2$) by less average ionic radii of *Na/Sm*. Diffuse type phase transition behavior was calculated by modified Curie-Weiss law and it was concluded due to random distribution of *Na/Sm* at the *Pb* site. These dielectric materials are important for technical applications.

# Acknowledgement

One of the authors Arun Kumar Yadav is grateful to the university Grants Commission to award me fellowship (NFO-2015-17-OBC-UTT-28455).Principle investigator expresses sincere thanks to Indian Institute of Technology, Indore for funding the research and also using the Sophisticated Instrument Centre (SIC) for the research.